\documentclass[12pt,a4paper]{article}
\usepackage[sumlimits]{amsmath}
\usepackage{amsfonts,amssymb,latexsym,eufrak,mathrsfs,graphicx,enumerate,bbm,bbold,epsfig}
\newcommand{\be}{\begin{equation}}
\newcommand{\ee}{\end{equation}}
\newcommand{\beq}{\begin{equation}}
\newcommand{\eeq}{\end{equation}}
\newcommand{\ba}{\begin{eqnarray}}
\newcommand{\ea}{\end{eqnarray}}
\newcommand{\bea}{\begin{eqnarray}}
\newcommand{\eea}{\end{eqnarray}}

\begin{document}
\baselineskip=15.5pt \pagestyle{plain} \setcounter{page}{1}


\def\del{{\partial}}
\def\vev#1{\left\langle #1 \right\rangle}
\def\cn{{\cal N}}
\def\co{{\cal O}}
\def\IC{{\mathbb C}}
\def\IR{{\mathbb R}}
\def\IZ{{\mathbb Z}}
\def\RP{{\bf RP}}
\def\CP{{\bf CP}}
\def\Poincare{{Poincar\`e}}
\def\tr{{\rm tr}}
\def\tp{{\tilde \Phi}}
\def\TL{\hfil$\displaystyle{##}$}
\def\TR{$\displaystyle{{}##}$\hfil}
\def\TC{\hfil$\displaystyle{##}$\hfil}
\def\TT{\hbox{##}}
\def\HLINE{\noalign{\vskip1\jot}\hline\noalign{\vskip1\jot}} 
\def\seqalign#1#2{\vcenter{\openup1\jot
  \halign{\strut #1\cr #2 \cr}}}
\def\lbldef#1#2{\expandafter\gdef\csname #1\endcsname {#2}}
\def\eqn#1#2{\lbldef{#1}{(\ref{#1})}%
\begin{equation} #2 \label{#1} \end{equation}}
\def\eqalign#1{\vcenter{\openup1\jot
    \halign{\strut\span\TL & \span\TR\cr #1 \cr
   }}}
\def\eno#1{(\ref{#1})}
\def\href#1#2{#2}
\def\half{{1 \over 2}}

\def\ads{{\it AdS}}
\def\adsp{{\it AdS}$_{p+2}$}
\def\cft{{\it CFT}}

\newcommand{\ber}{\begin{eqnarray}}
\newcommand{\eer}{\end{eqnarray}}
\newcommand{\beqar}{\begin{eqnarray}}
\newcommand{\cN}{{\cal N}}
\newcommand{\cO}{{\cal O}}
\newcommand{\cA}{{\cal A}}
\newcommand{\cT}{{\cal T}}
\newcommand{\cF}{{\cal F}}
\newcommand{\cC}{{\cal C}}
\newcommand{\cR}{{\cal R}}
\newcommand{\cW}{{\cal W}}
\newcommand{\eeqar}{\end{eqnarray}}
\newcommand{\eps}{\epsilon}
\newcommand{\pa}{\paragraph}
\newcommand{\pt}{\partial}
\newcommand{\de}{\delta}
\newcommand{\De}{\Delta}
\newcommand{\lb}{\label}


\newcommand{\oh}{\displaystyle{\frac{1}{2}}}
\newcommand{\dsl}
  {\kern.06em\hbox{\raise.15ex\hbox{$/$}\kern-.56em\hbox{$\partial$}}}
\newcommand{\id}{i\!\!\not\!\partial}
\newcommand{\as}{\not\!\! A}
\newcommand{\ps}{\not\! p}
\newcommand{\ks}{\not\! k}
\newcommand{\D}{{\cal{D}}}
\newcommand{\dv}{d^2x}
\newcommand{\Z}{{\cal Z}}
\newcommand{\N}{{\cal N}}
\newcommand{\Dsl}{\not\!\! D}
\newcommand{\Bsl}{\not\!\! B}
\newcommand{\Psl}{\not\!\! P}
\newcommand{\ZZ}{{\rm \kern 0.275em Z \kern -0.92em Z}\;}
\newcommand{\lbl}[1]{\label{eq:#1}}
\newcommand{ \rf}[1]{(\ref{eq:#1})}
\newcommand{\setl}{\setlength\arraycolsep{2pt}}

\newcommand{\noi}{\noindent}
\newcommand{\ra}{\rightarrow}
\newcommand{\Ra}{\Rightarrow}

\newcommand{\cd}{\bar{D}}
\newcommand{\cc}{\bar{C}}
\newcommand{\cB}{{\cal B}}
\newcommand{\cD}{{\cal D}}
\newcommand{\cG}{{\cal G}}
\newcommand{\cH}{{\cal H}}
\newcommand{\cK}{{\cal K}}
\newcommand{\cL}{{\cal L}}
\newcommand{\cM}{{\cal M}}
\newcommand{\cP}{{\cal P}}
\newcommand{\cS}{{\cal S}}
\newcommand{\cU}{{\cal U}}

\newcommand{\newc}{\newcommand}
\newc{\gsim}{\lower.7ex\hbox{$\;\stackrel{\textstyle>}{\sim}\;$}}
\newc{\lsim}{\lower.7ex\hbox{$\;\stackrel{\textstyle<}{\sim}\;$}}
\newc{\gev}{\,{\rm GeV}}
\newc{\mev}{\,{\rm MeV}}
\newc{\ev}{\,{\rm eV}}
\newc{\kev}{\,{\rm keV}}
\newc{\tev}{\,{\rm TeV}}
\newc{\ie}{{\it i.e.}}          \newc{\etal}{{\it et al.}}
\newc{\eg}{{\it e.g.}}          \newc{\etc}{{\it etc.}}
\newc{\cf}{{\it c.f.}}

\newcommand{\with}{\mbox{\rm with}}
\newcommand{\while}{\mbox{\rm while}}
\newcommand{\annd}{\mbox{\rm and}}
\newcommand{\foor}{\mbox{\rm for}}
\newcommand{\oll}{\mbox{\rm all}}
\newcommand{\att}{\mbox{\rm at}}
\newcommand{\are}{\mbox{\rm are}}
\newcommand{\hc}{\mbox{\rm h.c.}}
\newcommand{\too}{\mbox{\rm to}}

\input epsf

\begin{titlepage}

\leftline{OUTP-07-08-P}

\vskip -.8cm

\vskip 1.5 cm

\begin{center}

{\LARGE Warped Deformed Throats have Faster (Electroweak) Phase Transitions} \vskip .3cm

\vskip 1.cm

{\large {Babiker Hassanain\footnote{\tt babiker@thphys.ox.ac.uk},
John March-Russell\footnote{\tt jmr@thphys.ox.ac.uk}, and 
Martin Schvellinger\footnote{\tt martin@thphys.ox.ac.uk}}} 
\vskip 0.6cm

{\it The Rudolf Peierls Centre for Theoretical Physics, \\
Department of Physics, University of Oxford. \\ 1 Keble Road,
Oxford, OX1 3NP, UK.} \\

\vspace{1.7cm}

\begin{abstract}

We study the dynamics of the finite-temperature phase transition for warped Ran\-dall-Sundrum(RS)-like 
throat models related to the Klebanov-Tseytlin solution.   We find that, for infrared branes stabilized near the tip 
of the throat, the bounce action has a mild $N^2$ dependence, where $N(y) \sim  [M_5 L(y)]^{3/2}$ is the 
effective number of degrees of freedom of the holographic dual QFT, and where $L(y)$ is the local 
curvature radius, which decreases in the infrared.   In addition, the bounce action is not enhanced by large 
numbers.  These features allow the transition to successfully complete over a wider parameter range than 
for Goldberger-Wise stabilized RS models.  Due to the increase of $L(y)$ in the ultraviolet, the throat has a reliable 
gravitational description even when the number of infrared degrees of freedom is small.  We also comment on 
aspects of the thermal phase transition in Higgsless models, where the gauge symmetry breaking is achieved via 
boundary conditions. Such models include orbifold-GUT models and the Higgsless electroweak symmetry breaking 
theories of Csaki  {\it et al.}, with Standard Model gauge fields living in the bulk.

\end{abstract}

\end{center}

\noindent

\end{titlepage}

\setcounter{footnote}{0}

\newpage

\vfill

\section{Introduction}

The Randall-Sundrum (RS) model \cite{Randall:1999ee}, consisting of a warped
throat-like slice of AdS$_5$ space-time with ultraviolet (UV) and infrared (IR) 
boundary branes, offers a natural resolution to the electroweak hierarchy problem.  
Moreover, such strongly warped regions are a natural and possibly ubiquitous
feature of string theory flux compactifications \cite{warp,warp1,warp2,warp3,warp4,warp5,warp6,Hebecker:2006bn}, and,
due to the KKLT construction \cite{Kachru:2003aw}, the landscape of metastable string vacua.  
Therefore, it is a pressing question to understand the finite-temperature behaviour
of such theories and the possible implications of this behaviour for the early universe.

The finite-temperature equilibrium properties and phase-transition dynamics of
Gold\-ber\-ger-Wise (GW) \cite{Goldberger:1999uk} stabilized RS models  
have been studied in Refs.\cite{Creminelli:2001th,Randall:2006py,Nardini:2007me}. It was found that 
the RS solution is thermally preferred at temperatures much less than the IR scale,
while the high-temperature phase is an AdS$_5$-Schwarzschild
(AdS-S) geometry with a horizon replacing the infrared brane, with a first-order transition
between the two phases.  However, as noted in Ref.\cite{Creminelli:2001th}, essentially
because of the weak breaking of conformal symmetry in the IR for GW
stabilized RS models, the transition temperature is parametrically suppressed relative to the
IR (TeV) scale, and there is only a very small range of parameters where the transition is
able to complete, the early universe typically being stuck in a Guthian old-inflation state.
In addition, the part of parameter space where the transition successfully completes is the
regime in which a reliable gravitational description of the RS slice of AdS$_5$ is, at best, close
to breaking down.   These difficulties not only apply to the electroweak phase transition if the 
hierarchy problem is solved by warping, but they also have implications for hidden sector throats if 
the post-inflation reheat temperature in the hidden throat is above the IR scale of the throat.

Our intention in this letter is precisely to re-examine the issue of phase transitions in warped models.
In particular, we consider the more realistic class of warped throat solutions based on the 
Klebanov-Tseytlin (KT) geometry  \cite{Klebanov:2000nc,Klebanov:2000hb} that one expects in 
string theory constructions, instead of the rather idealized GW-stabilized RS models studied so far.   
Our primary result is that the thermal bubble
action has a much milder $N^2$ dependence (where $N$ measures the effective number of degrees of 
freedom of the dual holographic theory) and is not otherwise enhanced by large numbers, allowing the 
transition to successfully complete over a significantly wider range of parameter space.    In addition, an 
important feature of the KT geometry is that the effective $N$ increases with distance, $y$, along the throat 
as one moves away from the IR, $N^2(y) \simeq 27\pi^2 [M_5 L(y)]^3/4 g^2_{\rm str}$,
where $L(y)$ is the local curvature radius which becomes larger in the UV.  Because of this increase, 
and even for small $N$ in the IR, the majority of the throat can be in a regime where the gravitational 
description is good, $M_5 L(y) \gg 1$, and the existence of the strongly warped throat is reliably predicted 
in the first place.  

The plan of this letter is as follows:  in Section 2 we discuss the KT throat geometry as well as a
5D effective field theory description of the throat and its stabilization dynamics developed by
Brummer \etal\  \cite{Brummer:2005sh}.   In Section 3 we discuss the thermal
phase transition in the KT throat geometry, and compute the transition temperature as a function
of the warp factor at the tip of the throat.   Section 4 addresses the phase transition dynamics following the
procedure of Creminelli \etal\  \cite{Creminelli:2001th}, and also compares our results to those obtained
previously for the GW-stabilized RS geometry.   In Section 5 we discuss the nature of 
gauge symmetry breaking when gauge fields are present in the bulk of the throat.  
In particular, we focus on models such as Orbifold GUTs and Higgsless electroweak models, 
which involve gauge symmetry breaking by boundary conditions on the IR brane.  Finally, Section 
6 contains our conclusions.

\section{The Stabilized KT Throat \& the 5D Effective Theory}
\lbl{model}

The conifold throat region in the Klebanov-Tseytlin (KT) solution of type IIB supergravity is described by the 10D metric 
\be\label{KT}
ds^2 = h(r)^{-1/2} \eta_{\mu\nu} \, dx^\mu dx^\nu+ h(r)^{1/2} ( dr^2 + r^2 ds^2_{T^{1,1}} )  \, ,
\ee
with a constant dilaton, certain fluxes that we need not specify here, and 4D Minkowski metric
$\eta_{\mu\nu}$.  The warp factor is given by 
\be\label{KTwarp}
h(r) = 1 + \frac{81\, \alpha'^2 g_{\rm str}^2 M^2 \log(r/r_s)}{8 r^4} \, .
\ee
Here $M$ is the $F_3$-flux quantum (equivalently the number of fractional D3 branes at the conifold 
singularity) while $ds^2_{T^{1,1}}$ is the 5D metric of the internal $T^{1,1}$.  
The throat region, analogous to the region between the IR and UV branes 
in RS models, lies between $r/r_s \sim 1$ and 
$[81 \alpha'^2 g_{\rm str}^2 M^2 \log(r/r_s)]/(8r^4) \sim 1$ 
in the IR and UV, respectively. The fine 
details of the Calabi-Yau manifold onto which the throat matches at large $r$, or the region around $r=r_s$ 
(a singularity of the above metric which can be resolved by a Klebanov-Strassler tip \cite{Klebanov:2000hb}), 
will not play a role in our calculation.  What is important is that because of the $\log(r/r_s)$ in
Eq.(\ref{KTwarp})  this geometry describes a deformation away from an exact conformal $AdS_5\times \Sigma_5$,
 with the feature that the breaking of conformal symmetry becomes stronger as the IR is approached, 
 $r\rightarrow r_s$.   This is precisely the feature that will enable an unsuppressed thermal transition 
 \cite{Creminelli:2001th,Kaplan:2006yi}.

To study the dynamics of the thermal transition, it is useful to switch over to an effective 5D description
in which the essential degrees of freedom of the throat are isolated.   Brummer, Hebecker, and Trincherini 
(BHT)  \cite{Brummer:2005sh} have shown in detail how the flux stabilization of the throat length in the KT 
solution can be understood in terms of a Goldberger-Wise-like mechanism stabilizing a hierarchy between 
effective ultraviolet and infrared branes.  In the 5D Einstein frame the field content is gravity minimally 
coupled to a scalar field $H$ with action given at leading order by
\begin{eqnarray} \label{action}
{\cal {S}} &=&\int d^4x \int_{y_{IR}}^{y_{UV}} dy \sqrt{-G} \, \left(\frac{ M_5^3}{2} \mathcal{R}_5 -
\frac{1}{2} G^{MN}\partial_M H\partial_N H - V(H) \right) \,  \nonumber \\
                           &-& \int d^4x\, dy \frac{\sqrt{-G}}{\sqrt{G_{yy}}} \biggl([\Lambda_{IR}+V_{IR}(H) ]
                           \delta(y-y_{IR}) + [\Lambda_{UV}+V_{UV}(H) ]\delta(y-y_{UV}) \biggr) ,
\end{eqnarray} 
where $\mathcal{R}_5$ is the five-dimensional Einstein scalar curvature, $M_5$ is the Planck mass in 5D,
and $V$ and $V_{UV,IR}$ are the bulk and UV/IR-brane-localized potentials of the scalar $H$.  The new radial
coordinate $y$ is related to $r$ by
\be\label{ydef}
y = \frac{(3 g_{\rm str}^2 M^2/\pi^2)^{2/3}}{5 M_5} \, [\log(r/r_s)]^{5/3} \equiv R_s \, [\log(r/r_s)]^{5/3}  \, ,
\ee
and runs from small values, $y\sim R_s$ (which sets the size of the tip region) in the IR, to large values in the UV.   The potential is
\be\label{Hpot}
V(H) = -\frac{864 M_5^7}{25 R_s^2} H^{-8/3} \, .
\ee
The GW field $H$ represents the continuously varying (with respect to $r$ or $y$) flux of the Neveu-Schwartz 2-form potential 
$B_2$ integrated on the two-cycle in $T^{1,1}$, which deforms the geometry along the throat away from $AdS_5$.   Finally, the 
brane-localized tensions $\Lambda_{IR,UV}$ take account of the leading contributions of the
UV ``CY-head" and IR ``tip" in the 5D effective theory and are necessary for the satisfaction of the Israel boundary conditions.  

It is straightforward to see that the system of Eq.(\ref{action}) possesses solutions reproducing
the 5D $(x^\mu,r)$-coordinate part of the KT throat in Eq.(\ref{KT}).   Explicitly, consider the
5D metric ansatz 
\be\label{broken}
ds^2 = e^{2A(y)-2A(y_{UV})} \eta_{\mu\nu} \, dx^\mu dx^\nu+ dy^2  \, . 
\ee
The action of Eq.(\ref{action}) then implies the $H$ equation of motion 
$(\partial_y^2 + 4 \partial_y A(y) \partial_y) H = \partial V/\partial H$. 
{\it A posteriori}, it can be checked that, away from the far IR region, 
$H$ is slowly varying on the scale of the curvature, so the $\partial_y^2 H$ term can be dropped. 
The potential Eq.(\ref{Hpot}) then 
implies that the profiles of the warp factor $A$ and scalar $H$ are given by 
\be\label{AHsolns}
A=(y/R_s)^{3/5} + subleading,  ~~~{\rm and}~~~ H= \sqrt{8} M_5^{3/2} (y/R_s)^{3/10} + subleading \, ,
\ee
in an expansion in powers of $(R_s/y)$, or equivalently $1/A(y)$.  Examining the higher-order terms in the $1/A(y)$-expansion of 
the equations of motion shows that the expansion breaks down for $A(y) \lsim 1$, where the local curvature
length $L(y)$ defined by  $\mathcal {R}_5 = - 20/L^2(y)$ is given by
\be\label{localcurv}
L(y) \simeq \frac{5 R_s A(y)^{2/3}}{3} \, ,
\ee
and approaches the size of the ignored $T^{1,1}$ manifold.  For $A(y)\gsim 1$ one is in a regime where the 5D 
gravitational description is reliable.  Related to this fact, the number of degrees of freedom of the 
holographic dual gauge theory is large in this regime.  Specifically, at a given $y$ the number of effective degrees of 
freedom of the dual (${\cal N} =1$ SYM) gauge theory is given by
\be\label{KTdof}
N^2(y) = \frac{27 \pi^2}{4 g^2_{\rm str}} \, [M_5 \, L(y)]^3 \, ,
\ee
as can be deduced from the analysis of KT \cite{Klebanov:2000nc} and BHT \cite{Brummer:2005sh}. Note that in
the remainder of this work we take $g^2_{\rm str} = 1$, following Klebanov and Tseytlin \cite{Klebanov:2000nc}.

The solution Eqs.(\ref{broken},\ref{AHsolns}) reproduces the 5D geometry of the deformed warped throat of the 
KT solution Eqs.(\ref{KT},\ref{KTwarp}) after performing a Weyl re-scaling, as can be seen using Eq.(\ref{ydef}).
It is straightforward to derive from the solution the expression for the 4D Planck mass $M_{{\rm pl},4}$ in terms
of the parameters in the 5D effective theory:
\be\label{4DPlanck}
M_{{\rm pl},4}^2 = \frac{5}{6} M_5^3 R_s \left(\frac{y_{UV}}{R_s}\right)^{2/5} + subleading\, .
\ee

Finally, in the effective 5D theory, the positions of the IR and UV branes, and thus the relative warping 
between the IR and UV, are set by the values of $H$ enforced by the brane-localized potentials
$V_{IR,UV}$.    If the potentials are stiff, enforcing $H(y_{IR,UV})/(\sqrt{8} M_5) = c_{IR,UV}$, for some ${\cal O}(1)$ 
constants $c_{IR}$ and $c_{UV}$ respectively, then the relative warping between the ends of the throat is simply given by
$\exp\left(2A_{IR} - 2A_{UV}\right) = \exp\left(2c_{IR}^2 - 2c_{UV}^2 \right)$, and can easily be sufficiently large
to explain the electroweak-to-Planck hierarchy.  

Before we turn to the study of the thermal phase transition we comment
that we will not, at any point, require the exact form of the potentials $V(H)$ and $V_{IR,UV}$ or
the effective brane cosmological constants $\Lambda_{IR,UV}$.
The reason is that once the Einstein equations are
solved under the assumption of low curvature, everything in the 
problem is determined in terms of the warping $A(y)$. The 
relevant equations enabling this simplification are 
\begin{eqnarray}
6A^{'2} &\simeq& -\frac{V}{M_5^3}  \, , \\
\quad M_5^3\sqrt{-G}(3 A^{''}) &\simeq& -\sqrt{-G_{IR}} \, \Lambda_{IR} \, \delta(y-y_{IR}) - \sqrt{-G_{UV}} \, 
\Lambda_{UV} \, \delta(y-y_{UV}) .
\end{eqnarray}
In particular, the last equation is nothing but the expression of the jump condition for the
metric at the branes, and it gives us $\Lambda_{UV}=\frac{18}{5 R_s}M_5^3 A_{UV}^{-2/3}$ and 
$\Lambda_{IR}=-\frac{18}{5 R_s}M_5^3 A_{IR}^{-2/3}$. 

\section{The Thermal Phase Transition}
\lbl{phasetransition}

Following the analysis of Creminelli \etal\ \cite{Creminelli:2001th}, the features of the
finite-temperature phase transition are most easily explained by utilizing the holographic
dual description of the theory.  As is well-known, the AdS/CFT correspondence
\cite{Maldacena:1997re,Witten:1998qj, Gubser:1998bc} admits an extension to
RS models such that they possess a dual interpretation as a strongly coupled 4D CFT
coupled to gravity and spontaneously broken in the IR
\cite{Gubser:1999vj,ArkaniHamed:2000ds,Rattazzi:2000hs,Hebecker:2001nv}).   The further extension
to perturbed AdS theories, and thus non-CFT dual theories is also by now well understood (see \eg\
\cite{Zaffaroni:2005ty}).

We start by briefly recalling the gravity/QFT holographic dictionary: an operator $\hat{\mathcal{O}}$ on
the field theory side is sourced by the boundary value $\phi_0$ of a bulk field $\phi$ defined in the 
gravity theory. Mathematically: 
\begin{equation}
\left<\textrm{exp}{\int \mathrm{d}^4x \, \hat{\mathcal{O}}\phi_0} \right>_{QFT} 
= Z_{5D}[\phi_0] \, ,
\end{equation}
where the left hand side of this equation refers to the generating functional of the given 
boundary operator $\hat{\mathcal{O}}$, and the right hand side refers to the partition function 
calculated in the gravity theory, with the restriction of the field $\phi$ to the value $\phi_0$ 
on the boundary of the space. Using this ansatz, one may calculate the finite temperature 
partition function, and thus the thermodynamic properties, of the 4D field theory using the 
5D path integral with periodically identified Euclidean time. The advantage offered by the
5D formulation is that when the boundary theory is strongly coupled, the bulk theory is
weakly coupled, allowing the use of standard  semi-classical techniques.

Now, if we have a certain QFT defined on the boundary of the space, and we claim that this 
QFT is dual to a certain gravitational description, then there is no reason to pick one
solution to the bulk Einstein equations over another. To put it differently, any 5D bulk metric which solves 
the bulk Einstein equations and asymptotes to the required behaviour at the UV is in principle 
admissible in the holographic correspondence, the different metrics 
corresponding to different (thermal) phases of the dual 4D field theory. The preferred phase of 
the QFT is found by comparing the free energies of the different gravitational backgrounds
\cite{Witten:1998zw}. 

Thus, to study the phase transition in our system we first demonstrate an alternative solution to the 
bulk Einstein equations, namely a warped black hole solution. 

\subsection{The black hole solution}

Examining the Einstein equations coming from Eq.(\ref{action}), and proposing a black hole ansatz of the form
\be
ds^2 = e^{2A(y)-2A(y_{UV})} \left( -f(y) dt^2+\delta_{ij} \, dx^i dx^j \right)+ \frac{dy^2}{f(y)}  \, ,
\ee
where $i,j=1,2,3$,
we find that the equations are indeed solvable, again subject to the assumption of $A(y)\gg 1$.
The function $f(y)$ is given by an integral expression in terms of the warping $A$, so that 
\be
f(y)=1-\frac{\int^y dy'e^{-4A(y')}}{\int^{y_h} dy'e^{-4A(y')}} \simeq 1-\left(\frac{A(y)}{A_h}\right)^{2/3} e^{4A_h-4A(y)} ,
\ee
where $A_h=A(y=y_h)$.
This solution has a black hole horizon at $y=y_h$, and the IR brane is eliminated. The UV brane remains 
with the same brane tension as before, because of the essential requirement of holography, which is that
the induced metric at the UV is identical for any bulk configuration. 

However, the profile of the field $H$ in this background is different, meaning that the potential $V$ as a function of 
$y$ is different.  Again, we can read off the required dependence from the Einstein equations, obtaining
\begin{equation}
6 f A'^{2}+\frac{3}{2}A' f' \simeq -\frac{V}{M_5^3} .
\end{equation}
This will be crucial in determining the properties of the phase transition. Note also that the temperature of this black hole, 
which can be calculated by ensuring that there is no conical singularity at the horizon, is given at leading order
in the expansion in $1/A_h$ by 
\be\label{horizontemp}
T_h=\frac{3A_h^{-2/3}e^{A_h-A_{UV}}}{5\pi R_s} .
\ee
As alluded to above, the black hole solution is thermally preferred at high temperatures. 
In this case, there is a black hole horizon instead of the infrared brane. At the position of the 
UV brane, which is far from the black hole horizon, the geometry is similar to that of Eq.(\ref{broken})
We now show that this solution of the Einstein equations is thermodynamically stable above a critical temperature $T_c$. 

\subsection{Comparing the free energy}

We wish to calculate the free energy of both gravitational solutions. To do this, we need to Wick-rotate the time 
coordinate to go into Euclidean space, such that the time is periodic with periodicity $1/T$, where $T$ is the 
temperature of the ensemble. We must then employ the semi-classical approximation, calculating the 
on-shell action $S$. The free energy is then given by $-T \, S$. 

This amounts to evaluating $\mathcal{R}_5$ via the Einstein equations and substituting into Eq.(\ref{action}) for both 
gravitational solutions. Taking the trace over the Einstein equations, one obtains
\begin{eqnarray}
\frac{1}{2}M_5^3 \, \sqrt{-G} \, \mathcal{R}_5 &=& \frac{1}{2}\sqrt{-G} \, G^{yy} \, (\partial_yH)^2+\frac{5}{3}\sqrt{-G} \, V(H) \nonumber  \\ 
            &+&  \frac{4}{3}\frac{\sqrt{-G}}{\sqrt{G_{yy}}} \, \Lambda_{UV} \, \delta(y-y_{UV})+\frac{4}{3}\frac{\sqrt{-G}}{\sqrt{G_{yy}}} \, \Lambda_{IR} \, \delta(y-y_{IR}) .
\end{eqnarray}
Note that this is the form of the equation in either gravitational background, assuming that $H$ depends only on $y$. 
The $y$-dependence of $V$ is of course different in either background, as argued above. Plugging this into 
the action for the broken phase (\ie, the warped solution Eqs.(\ref{broken},\ref{AHsolns}) with both UV and IR branes),
we obtain
\begin{equation}\label{potential}
S_b=-\frac{M_5^3}{10R_s}\beta \left[  A^{-5/3}e^{4A-4A_{UV}}\right]_{IR}^{UV} +\mathcal{O}(A_{UV}^{-8/3}) ,
\end{equation}
where this is the on-shell action per unit 3-volume. Note that, to be consistent with the large $A$ approximation, 
we have kept only the leading terms of order $A^{-5/3}$ and higher. The parameter $\beta$ is the inverse of the 
temperature, and comes about when the integral over Euclidean time is performed. The free energy of this phase is 
given by $F_b=-S_b/\beta$. 

Carrying out the same calculation for the black hole solution (unbroken phase), we obtain 
\begin{equation}
S_u=\frac{M_5^3}{10R_s}\beta \left( [4 \ln (A_{UV}/{A_h})+3]
A_h^{-2/3}e^{4A_h-4A_{UV}}-\left[ A^{-5/3}e^{4A-4A_{UV}} \right]_h^{UV} \right)+\mathcal{O}(A_{UV}^{-8/3}) ,
\end{equation}
where the free energy of the black hole is given by $F_u=-S_u/\beta$.

One immediately sees that the parts of the free energies which have a power-law dependence on the ultraviolet physics vanish 
identically upon subtracting the two free energies, leaving a mild logarithmic residual dependence on $A_{UV}$. The critical  
temperature $T_c$ of the transition can be obtained by equating the two free energies, giving the equation
\begin{equation}\label{AHAIReqn}
 [4 \ln (A_{UV}/A_h)+3] A_h^{-2/3}e^{4A_h-4A_{UV}} \simeq A_{IR}^{-5/3}e^{4A_{IR}-4A_{UV}} ,
\end{equation}
where we have used only the leading terms.  Note that from this equation we find
$A_h < A_{IR}$.    This is an important feature as it implies that the thermal transition occurs
in a region where conformal symmetry breaking is greater than that for
the position of the IR brane at zero-temperature.  
Although the equation for $A_h$ can not be solved analytically, an approximate solution,
valid for values of $A_{IR}\gsim 1.5$, is
\begin{eqnarray}
A_h &\simeq & A_{IR}-\frac{5}{12}\ln(A_{IR}) - \frac{1}{4} \ln[4 \ln(A_{IR}+\Delta)+3] \nonumber \\ 
    &+& \frac{1}{6}\ln\left[A_{IR}-\frac{1}{4}\ln\left( A_{IR}^{5/3}[4\ln(A_{IR}+\Delta)+3 \right) \right] .
\end{eqnarray}

Rewriting the free energy of the unbroken phase in terms of the temperature, 
and rewriting the free energy of the broken phase to expose the dimensions, we obtain
\begin{eqnarray}
F_b &\simeq& -\frac{(M_5 R_s)^3}{10}A_{IR}^{-5/3}\left( \frac{e^{-\Delta}}{R_s} \right)^4 \quad \textrm{and}  \\
F_u &\simeq& -\frac{(M_5 R_s)^3}{10} \left( (5\pi/3)^4 \alpha[\ln(5 \pi R_s T_c/3)+A_{UV}]^2 \right) T_c^4   ,
\end{eqnarray}
where $\alpha=4 \ln [A_{UV}/A_h(T_c)]+3$, and $\Delta=A_{UV}-A_{IR}$. This allows us to express
the temperature of the transition as
\begin{equation}\label{Tc}
T_c=\frac{A_{IR}^{-5/12}}{\tilde{\alpha}^{1/4}} \left( \frac{e^{-\Delta}}{R_s} \right) \, ,
\end{equation}
where, after some algebra,
\begin{equation}\label{alphaeqn}
\tilde{\alpha}(A_{IR})^{1/4}=\frac{5\pi}{3} \left( 4 \ln [A_{UV}/A_h]+3\right)^{1/4}  A_h^{1/2} .
\end{equation}
In this last equation $A_h$ should be understood as the $A_{IR}$-dependent solution to Eq.(\ref{AHAIReqn}).
This shows us that there is no hidden $N$-dependence inside $\tilde{\alpha}(T_c)^{1/4}$, although there is
$A_{IR}$ dependence.  Notice that the factor $\tilde{\alpha}(T_c)^{1/4}\leq 10$ for $A_{IR}\leq 2$.  

For a throat that solves the electroweak-to-Planck hierarchy problem by warping, $e^{-\Delta}/R_s$ is 
of the order of a TeV.  This then immediately leads to the observation that $T_c$ is  
also ${\cal O}(\tev)$ and is not parametrically suppressed if $A_{IR} \sim 1$, \ie\ if the IR brane is in the tip 
region as one expects.  This is in contrast to the result of Creminelli \etal\  Ref.\cite{Creminelli:2001th},
who find that the transition temperature is $T_c = (8\epsilon^{3/2} v_1^2/ \pi^2 N^2)^{1/4} \mu_{\tev}$, which is
suppressed relative to $\mu_{\tev}$ by both a power of $N$ (where $N^2 \simeq 16\pi^2 [\tilde{M}_5 L]^3$ 
measures the number of degrees of freedom of the dual QFT in their case 
\footnote{Note that the mass scale $\tilde{M}_5$ is related to our $M_5$ by the rescaling $M_5^3=4 \tilde{M}_5^3$, 
due to a factor four difference in the gravitational action.}) and the additional small parameters
$v_1$ and $\epsilon$ related to the GW stabilization mechanism (typical values are
$\tilde{M}_5 L\gsim 5$, $\epsilon\simeq 1/20$ and $v_1 \lsim 1/5$).

\section{The Dynamics of the Phase Transition}
\lbl{dynamics}

We have shown that there are two gravitational solutions corresponding to two different finite-temperature
phases of the theory. There is a free-energy barrier between the two phases at $T_c$ so the
thermal transition is first order, proceeding via the nucleation of critical bubbles of the stable phase inside 
the unstable phase, which then grow eating up the false phase.  If the nucleation rate per Hubble volume,
$\Gamma/H_{\rm Hub}^3$, is higher than the rate of expansion of the universe, $H_{\rm Hub}$, the bubbles
of stable phase will collide, and the phase transition will complete.  

To calculate the rate of bubble nucleation, the rigorous procedure is to construct the full gravitational (and stabilization)
field dependent `bounce' solution corresponding to the bubble nucleation, and then calculate the on-shell action for this
solution, and ideally, also calculate the fluctuation determinant about this solution \cite{Coleman:1977py, Callan:1977pt}.  
Unfortunately, the precise nature of the
topology-changing gravitational instanton is not known so we cannot follow this procedure.  However, following the analysis of 
Creminelli \etal\ and Randall and Servant we are able to estimate the on-shell bounce action at leading order in a large $A_{IR}$
expansion, which is sufficient for our purposes.  These authors imagine a configuration which interpolates between the black 
hole solution outside, going through pure AdS, then pushing the infrared brane from infinity (far IR) back to its stabilized position.
There are then two contributions to the bounce action, one from motion in the broken phase, and one from the unbroken phase.
Consider the broken phase contribution:  the degree of freedom whose motion dominates this part of the bounce action
is the massive radion field which encodes the inter-brane separation.  To calculate the bounce action we need to know
the form of the radion free energy as a function of the interbrane separation, and, in the case of GW-stabilized RS solutions, this is known to leading order.   In our case there is similarly a massive radion in the broken phase.  This radion field is stabilized 
at a certain value via the fluxes at either end of the space, and the value of the radion free energy function at its minimum is exactly 
the on-shell action in the broken phase that we have calculated.  Moreover, as we argue below, we know both the physical height 
and length of the free-energy barrier on the broken side, although not the full functional form.  To be conservative, we thus assume 
that the potential for the radion field to tunnel from small values to its stabilized value is a square potential barrier.  
This should give us a worst-case estimate for the contribution to the tunneling rate only from the broken phase side of the geometry.

The contributions from the black hole side, as well as the small $y/R_s$ region of the broken side,
are not calculable, similar to the case of the analysis of Creminelli \etal\ and Randall and Servant.  However,
because these contributions arise from a strong coupling region with dynamical scale $\Lambda$ we can estimate their
contribution to the free-energy barrier and thus the tunneling action.  As we will argue in the next section, 
this contribution is sub-leading
in the large $A_{IR}$ limit.

\subsection{The radion kinetic energy term}
 
By the arguments we just gave, we already have the height of the free energy barrier between the two phases, which is given 
by Eq.(\ref{potential}).  What we also need, however, is the canonical kinetic energy term for the radion in the broken phase, 
so that we can evaluate the correctly normalized tunneling distance from the $A\sim 1$ regime to the stabilized radion value. 
So the question is: how do we represent the radion in our metric of Eq.(\ref{broken})?  We use the simplest available ansatz 
to achieve that, namely requiring that the metric representation still solves the Einstein equations for a constant value of the 
radion.  The ansatz we employ is \cite{Goldberger:1999un,Csaki:1999mp,Charmousis:1999rg}
\be\label{radion}
ds^2 = e^{\left[ 2A(y)-2A(y_{UV})\right] \phi(x)^{3/5}} g_{\mu\nu}(x) \, dx^\mu dx^\nu+\phi(x)^2 dy^2  \, ,
\ee
where $\phi(x)$ is the as yet un-normalized radion field. Note that we have chosen conventions where the minimum of the
radion is at $\phi=1$.  Now, to obtain the kinetic energy term in the dimensionally reduced effective 4D action, we
need to only retain terms with two powers of the radion  and two powers of the derivative in the $x^{\mu}$ direction. We finally obtain
\begin{equation}
S_{KE}=-\frac{9M_5^3 R_s}{10}\int\mathrm{d}^4x\sqrt{-g(x)} \, A_{IR}^{2/3} \, \Delta^2 \, e^{-2\Delta \phi^{3/5}} \, \phi^{-2/5} \, 
g^{\mu\nu} \, \partial_\mu \phi \, \partial_\nu \phi  .
\end{equation} 
Carrying out an (approximate) normalization for this term, we obtain that the tunneling distance is
from about zero to $\chi_{min}$ in the normalized coordinates $\chi$, where the physical tunneling
distance in field space is given by
\begin{equation}\label{distance}
\chi_{min}=\sqrt{5}\,  (M_5 R_s)^{3/2} \,  A_{IR}^{1/3} \, \left( \frac{e^{-\Delta}}{R_s} \right)  .
\end{equation}
This tunneling distance is of the same order as that of Creminelli \etal. This is to be expected: the radion is a gravitational degree 
of freedom, and therefore we expect the factor of $(M_5R_s)^3$ sitting outside its kinetic term, in the same way as for the 4D 
graviton.  By comparing this tunneling distance with the critical temperature of Eq.(\ref{Tc}), we can see that the tunneling distance 
is bigger by a factor
\begin{equation}\label{largechi}
\frac{\chi_{min}}{T_c}\sim (M_5 R_s)^{3/2} A_{IR}^{5/4} \sim N_{IR}  A_{IR}^{1/4}  \, ,
\end{equation}
where $N_{IR}=N(y_{IR})$.  Because of this
the contribution to the bounce action from the black hole side of the instanton is parametrically suppressed 
(at least until $M_5 R_s, A_{IR} \sim 1$) compared to the contribution from the broken side.  This is
the same as the analysis of Creminelli {\it et al}, and allows us to focus on the broken-side contribution.
At this point we have all the information needed to calculate the rate of bubble nucleation, to
which we now turn.
 
\subsection{The rate of bubble nucleation} 

We will now collect all of intermediate results and assemble them into the rate of bubble nucleation of broken phase bubbles inside 
the black hole phase.  As in the case discussed by Randall and Servant \cite{Randall:2006py}, the transition typically proceeds 
by $O(3)$-symmetric thermal nucleation of thick-walled bubbles.  We will therefore quote the (approximate) formulae for thermal 
bubble nucleation in the thick-wall approximation, and then use our results to do the calculation. 

For thermal bubbles, the formula for the rate of bubble nucleation per unit 3-volume is 
\begin{equation}\label{condition}
\Gamma=B e^{-S_3/T} .
\end{equation}
The fluctuation determinant $B\simeq T_c^4$ at the transition temperature, and $S_3$ is the Euclidean action for the 
bounce solution, which is given by\footnote{Note that there is a typographic error in the $T$ dependence of Eq.(18) of Ref.\cite{Randall:2006py}.}  
\begin{equation}\label{bubble}
S_3 \simeq \frac{4\pi}{3} \frac{\chi_{min}^3}{\sqrt{2 \, \delta F_T}}  \, ,
\end{equation}
where, as we explained above, we have made the conservative assumption of taking the potential for the radion, $V(\chi)$, to
be just a square barrier with length $\chi_{min}$. The denominator in Eq.(\ref{bubble}) is the difference in
the free energies of the phases at temperature $T$, which we can write in the form
\begin{equation}
\delta F_T=\frac{(M_5 R_s)^3}{10}A_{IR}^{-5/3}\left(\frac{e^{-\Delta}}{R_s}\right)^4 \left[1-\left(\frac{T}{T_c} \right)^4
\frac{\tilde{\alpha}(T)}{\tilde{\alpha}(T_c)} \right].
\end{equation}
Gathering all the ingredients, we now have an expression for $S_3/T$, in which we display clearly the 
dependence of the tunneling rate on the relevant parameters.  After using the expression for $T_c$,
Eq.(\ref{Tc}), we find
\begin{equation}\label{S3}
\frac{S_3}{T}=\frac{100\pi}{3} \tilde{\alpha}(T_c)^{1/4} (M_5 R_s)^3  A_{IR}^{9/4} \,  f(T)  \, ,
\end{equation}
where
\begin{equation}\label{fT}
f(T) = \frac{T_c}{T} {\left(1-\left(\frac{T}{T_c}\right)^4\frac{\tilde{\alpha}(T)}{\tilde{\alpha}(T_c)}\right)^{-1/2}} \, .
\end{equation}
Keeping in mind that $\tilde{\alpha}(T_c)^{1/4}$ has a leading $A_{IR}$ dependence of $A_{IR}^{1/2}$, it is clear that
Eq.(\ref{S3}) demonstrates the strong suppression of the bounce action as one moves the position of the infrared
brane closer to the strong coupling regime, $A_{IR}\rightarrow 1$.   Of course as one approaches
$A_{IR} \sim 1$, we lose control over our calculation.   Nevertheless, we believe that Eq.(\ref{S3}) reliably shows
that the tunneling rate becomes parametrically unsuppressed as $A_{IR}$ becomes small.

Alternatively, in terms of the number of effective degrees of freedom of the holographic dual, Eq.(\ref{KTdof}),
we can write 
\begin{equation}\label{S3eq}
\frac{S_3}{T}=\frac{16}{15\pi} \tilde{\alpha}(T_c)^{1/4} A_{IR}^{1/4} N_{IR}^2 \,  f(T) \, .
\end{equation}
This is the form in which the constraints imposed by successful completion of the transition are physically most
transparent.  Examining this formula, we see that the bounce action goes like $N^2$, as expected from general 
considerations in strongly coupled theories \cite{Kaplan:2006yi}.   This is a major difference from the case of traditional
Goldberger-Wise stabilized Randall-Sundrum models, where a dependence of $N^{7/2}$ is obtained 
\cite{Creminelli:2001th,Randall:2006py}.   

As we have already mentioned, the criteria for the completion of the transition is that the bubble nucleation rate must 
exceed the rate of expansion of the universe, so that the bubbles collide. We can recast this condition as
\begin{equation}
\Gamma \gsim H_{\rm Hubble}^4 \quad \mathrm{where} \quad H_{\rm Hubble}^4 \sim \rho^2/M_{pl}^4  \, ,
\end{equation}
where $H_{\rm Hubble}$ is the Hubble parameter, $\rho$ is the energy density of the universe and $M_{pl}$ is the 
reduced 4D Planck mass. For phase transitions occurring at the electroweak scale, this implies that, upon taking the logarithm
of Eq.(\ref{condition}), the condition $S_3/T \lsim 140$.   Taking, in the expression Eq.(\ref{S3eq}) for the bounce action,
$A_{IR}=2$ (so $\tilde{\alpha}(T_c)^{1/4} \simeq 10$), and using the fact that the 
minimum value of $f(T)$ is $f\simeq 1.6$, we therefore obtain the following upper bound on $N_{IR}$
\begin{equation} \label{NconditionTEV}
N_{IR}^2 \leq 21  \, . 
\end{equation}
This restriction on $N_{IR}$ is stringent but materially less so than in the case of GW-stabilized RS models. 
We also remind the reader that Eq.(\ref{NconditionTEV})
is a conservative estimate, because we have taken the extreme approximation of the radion
potential as a square barrier of height equal to the maximum depth of the free-energy potential.

\subsection{The rate of bubble nucleation at other energy scales}

We have shown in the previous subsection that, for a throat in which the electroweak-to-Planck hierarchy
is resolved, the upper bound on $N_{IR}$ is improved compared to the result in Goldberger-Wise-stabilized 
Randall-Sundrum models. 
One can use the same formalism to tackle other hierarchies in field theory, the most obvious example
being the GUT-to-Planck hierarchy. The difference in that case would be the value of $\Delta$, which
measures the separation between the UV and IR branes.  More precisely, we now require that $e^{-\Delta}/R_s$
is of the same order as the GUT scale. 

The action for thick-walled bubbles still has the same functional form as Eq.(\ref{S3eq}), and the slow variation
of $\tilde{\alpha}(T_c)^{1/4}$ means that the actual rate of bubble nucleation does not change.
Applying the same reasoning as above, we obtain the following upper bound on $N_{IR}$
\begin{equation} \label{NconditionGUT}
N_{IR}^2 \leq 3  \, . 
\end{equation}

One can generally see that as we increase the scale at which the phase transition takes place,
the upper bound on $N_{IR}$ considerably decreases, meaning that the gravitational description
in the tip region is not trustworthy if the transition is to complete.

\subsection{Discussion of results}

The calculations of the bubble nucleation rate of Creminelli \etal\  in Ref.\cite{Creminelli:2001th}
have been done in the thin-wall approximation, where the radius of the bubbles is much larger
than the wall thickness, and for positive values of the GW parameter $\epsilon$.
More recently, Randall and Servant \cite{Randall:2006py} have studied the thick wall case (and $\epsilon < 0$),
where the wall thickness is comparable to bubble size, and they argued that this
contribution dominates and enhances the nucleation rate over most of parameter space.
However, for both Creminelli \etal\ and Randall and Servant, the bounce action still has a
strong $N^{7/2}$-dependence, so there is only a small range of parameter space
where the transition successfully completes.  Moreover, in the GW-stabilized RS case, the
expression $N^2 = 16 \pi^2 [\tilde{M}_5 L]^3 +1$ applies at all positions along the $AdS_5$ slice, so to be
in a regime where the gravitational description is under control one needs $\tilde{M}_5 L \gsim 1$ at the very least,
or equivalently $N\gsim 4\pi$,  which poses a significant problem for the completion of the phase transition.  

On the other hand, for the KT geometry, and unlike the GW-stabilized RS case, the conformality of the theory
while good in the UV is badly broken in the IR.   This can be seen explicitly from the fact that the number
of degrees of freedom of the dual gauge theory varies as a function of scale, Eq.(\ref{KTdof}),
$N^2(y) \simeq [M_5 L(y)]^3$, where $L(y)$ varies as
$L(y)\sim R_s (y/R_s)^{2/5}$ and therefore becomes small in the
IR region $y\lsim R_s$, and the t'Hooft parameter of the corresponding holographically dual QFT
becomes ${\cal O}(1)$, so the theory is truly strongly coupled.
Further, it is vital to realize that the $N^2$ in the bounce action is the IR value
of the number of degrees of freedom, and it is only this number that is
constrained to be small by completion of the phase transition, Eq.(\ref{NconditionTEV}).
In most of the throat the number of effective degrees of freedom, or equivalently the curvature radius, is much
larger so that the throat for most of its length is in a regime where the leading order action, Eq.(\ref{action}), applies,
and the existence of the throat is reliably described.

Finally, we mention in passing that transitions of {\it second-order} sometimes seem possible in the modified
geometries that we discussed in this paper.  We hope to return to this issue in future work.

\section{Gauge Symmetry Breaking and Higgsless models}
\lbl{higgsless}

So far, we have imagined that a physical Higgs degree of freedom is localized on
the infrared brane, with the electroweak gauge bosons living on either the IR brane or propagating
in the bulk of the space.  As argued by Creminelli {\it et al},  Standard Model fields localized on the
IR brane do not greatly affect the transition temperature or dynamics, as the contributions of
these modes to the free energy is formally sub-leading in the $N$-expansion, though 
one can imagine situations where this is no longer the case.   The SM gauge fields, whether they are
IR-localized or in the bulk, also do not greatly affect the transition, because they do not possess a tree-level 
contribution to the free energy in either case.  

However, models with IR branes lead to an interesting possibility regarding the breaking of 
electroweak symmetry.   In the case of bulk gauge fields, it is possible to completely remove
the Higgs from the problem:  by imposing a judicious set of boundary conditions on the gauge
fields, the zero modes of the gauge bosons can be removed, and one obtains a spectrum
of first-excited-state Kaluza-Klein modes which reproduce the spectrum of massive gauge bosons 
expected of a spontaneously broken gauge symmetry, but without the need for an 
explicit Higgs field.
\footnote{Of course, one can always consider this ``Higgsless'' case
to be a limiting case of the theory where the Higgs is localized on the 
IR brane, with the Higgs mass taken to the cutoff. But the point is that the
massive gauge boson scattering amplitudes are not unitarized by a light Higgs, but
rather by the spectrum of massive Kaluza-Klein gauge boson modes.} 
This gauge symmetry breaking by boundary conditions was
applied to extra-dimensional `orbifold' GUT model building in 
Refs.\cite{orbifoldGUTs,orbifoldGUTs1,orbifoldGUTs2,orbifoldGUTs3,orbifoldGUTs4}, and later
applied to electroweak symmetry breaking by Csaki \etal \cite{Csaki:2003dt,
Csaki:2003zu,Csaki:2003sh,Csaki:2005vy}.  This class of Higgsless electroweak models can be 
tweaked quite naturally in order to have properties very similar to those of the real world. 

It is interesting to discuss the manner in which the gauge symmetry is restored at high temperature in such
Higgsless theories.  As discussed in previous sections, the black hole solution is thermally preferred
at high temperatures, and in that case, the infrared brane is replaced by the black hole horizon. At the position of the 
UV brane, which is far from the black hole horizon, the UV boundary conditions are untouched. 
However, as we argue below one does not have any freedom in the infrared now: one is forced to select the regular 
solution of the equations of motion at the black hole horizon.  Since in the high temperature phase one
can only impose a regularity boundary condition on the black hole horizon, independent of the gauge index, 
the broken (electroweak or GUT) gauge symmetry is restored.

Let us show why this is the case. 
Consider the Euclidean equation of motion for the spatial directions of the gauge fields, in the gauge 
$A_5=0$, and with the added simplification of setting $A_t=0$ 
\begin{equation}
A_i^{''}+\frac{f'+2fA'}{f}A_i^{'}-\frac{q^2e^{-2\bar{A}}}{f}A_i-\frac{w^2e^{-2\bar{A}}}{f^2}A_i=0 \, ,
\end{equation}
where we have made a Fourier transformation in both the 4D spatial and temporal directions, denoting 
the spatial momentum by $q$ and the frequency by $w$. The dashes refer to differentiation with respect 
to $y$, and we have defined $\bar{A}(y)=A(y)-A_{UV}$. This equation has a regular singular point at 
the black hole horizon, as can easily be verified by expanding the equation near the singular point, 
obtaining
\begin{equation}
A_i^{''}+\frac{1}{x}A_i^{'}-\frac{q^2 e^{-2\bar{A}(y_h)}}{f'(y_h)x}A_i
-\frac{w^2 e^{-2\bar{A}(y_h)}}{f'^2(y_h)x^2}A_i=0 \, .
\end{equation}
The coordinate $x$ is given by $x=y-y_h$, so that the 
horizon is at $x=0$. From this equation, one can deduce the behaviour of $A_i$ near the horizon: 
the indicial equation gives the roots $\pm \bar{w}$, where $\bar{w}=e^{-\bar{A}(y_h)}w/f'(y_h)$ 
and so there is a regular solution and a linearly independent one which is 
divergent at the horizon.  We are thus forced to select the regular solution 
at the black hole horizon, as claimed. This is independent of the boundary conditions on the UV brane, 
which are unmodified.  Moreover, one can also see that for any gauge direction 
in which the UV boundary condition is Neumann, the regularity condition guarantees that the 
above equation can be solved for zero momentum. In other words, the masslessness of the zero modes 
of the gauge fields is restored in the black hole geometry as a direct consequence of regularity 
on the black hole horizon.  The zero mode profile is flat in the extra dimension, which obviously 
solves the above equation with $w=q=0$.

\section{Conclusions}
\lbl{conclusions}

In this letter we have re-examined the dynamics of the finite temperature (electroweak) phase transition in 
warped Randall-Sundrum-like throat models, in particular focusing upon a more realistic class of warped throat
solutions based upon the Klebanov-Tseytlin geometry.    This geometry is much
closer to the geometry one expects in typical string-derived warped throat constructions than the $AdS_5$ slice 
usually assumed.   For IR branes stabilized near the tip of a KT throat, we found that the transition rate is not 
parametrically suppressed beyond the expected $N^2$ dependence in the bounce action.  This enhancement 
in rate compared to that previously obtained in Goldberger-Wise stabilized RS models allows the transition to 
successfully complete over a wider range of parameter space.  Moreover, it is important that the $N^2$ in the 
bounce action is the IR value of the number of degrees of freedom, and it is only this number that is
constrained to be small by completion of the phase transition.   Because of the deformed warped nature of the
throat the number of effective degrees of freedom (or equivalently the curvature radius) at higher, UV, scales is
larger so that the throat for most of its length is in a regime where the leading order gravitational description is reliable
even if $N_{IR}\sim1$.  Finally, we also commented on aspects of the gauge symmetry breaking thermal phase 
transition in Higgsless models with boundary condition breaking, such as orbifold-GUT models and the Higgsless 
electroweak symmetry breaking theories of Csaki \etal\ with bulk Standard Model gauge fields, and showed 
precisely how the IR boundary conditions
implied by the horizon of the high-temperature black hole phase lead to gauge symmetry restoration.

~

~

\centerline{\bf Acknowledgments}

~

JMR is grateful to Raman Sundrum for discussions.  JMR acknowledges the kind hospitality of the Stanford Theory 
Group, and BH and JMR that of the CERN Theory Group during the course of this work.  This work has been supported 
by PPARC Grant PP/D00036X/1, and by the 'Quest for Unification' network, MRTN-CT-2004-503369.  The work of BH 
was supported by the Clarendon Fund, Balliol College and Christ Church College.

\end{document}